\documentstyle{article}

\title {On polynomial approximation of the static vortex 
in Abelian Higgs model}
\author {J. Karkowski 
	\thanks{Institute of Physics, Jagellonian University,
	30-064 Krak\'{o}w, Reymonta 4, Poland}
	\ and
	Z. \'{S}wierczy\'{n}ski
	\thanks{Institute of Physics, Pedagogical University, 
	Podchor\c a\.{z}ych 2, 30-084 Krak\'{o}w, Poland}}

\begin{document}
\maketitle

\begin {abstract}
The static vortex solution in Abelian Higgs model with small ratio
of vector and Higgs particle masses is considered. Several formulae
approximating  this solution are discussed. The accuracy of these
approximations is tested by numerical computations.
\end {abstract}

\newpage

\vspace{0.5cm}
\indent

\section{Introduction}
\indent

Nowadays vortex  solutions are found to be interesting in many areas 
of physics. Their investigations can help in better understanding 
of some phenomena in field theory, cosmology and condensed matter 
physics \cite{GEN}. It is rather difficult to obtain  vortex solutions
since one has usually to solve highly non-linear, very  complicated 
differential equations. Therefore appropriate analytical and numerical
methods must be worked out and applied to make some progress in this area.

One of the simplest systems possesing vortex solution is the Abelian
Higgs model. An exact vortex solution, in the form of infinite convergent
series, was found for this model in the
so-called  Bogomolny limit (the masses of the scalar and vector bosons
are the same)\cite{BOG}. In this limit the equations of motion reduce to the 
first order differential equations and the underlying methods can not
be simply applied in general case.
 
Recently in the Abelian Higgs model a polynomial approximation was 
extensively used to investigate excited vortex \cite{AH1},\cite{AH2}.
This method
gives analytical formulae simple enough to be applied in further 
computations. However it is necessary to estimate the error of this
approximation and this can be done by a comparison with the numerical
solution.

The goal of the present paper is twofold. First we would like to test
some approximating formulae for the static vortex solution proposed in
paper \cite{AH1}. These  analytic formulae were obtained in the limit
$ \kappa \to 0 $ where kappa is the ratio of vector and Higgs particle
masses. Our investigations are also limited to small values of $ \kappa $.
Second we would like to present more precise analytical approximations.
The accurate analytical and numerical results for the static vortex
solution can be useful in many problems apart from the excited vortex.
For example, equations governing evolution of a curved vortex \cite{A3}
involve  constants which are determined by the static vortex solution.

Our paper is organized as follows. In Sec. 2 we shortly review the simple
analytic formulae approximating the static vortex solution. As  
mentioned above these formulae were introduced and discussed in more 
detail in \cite{AH1}. We also compare the results obtained this way with 
numerical approximations to test the accuracy of the algorithm.
In Sec. 3 we propose several improvements of this method to get more 
precise analytical approximations and numerical results valid in wider
region than the previous ones. Finally in Sec. 4 we present some general
remarks and conclusions summarizing our paper. 


\section{ Simple approximations }
\indent

The Abelian Higgs model in $ 3+1 $-dimensional Minkowski space-time 
is described by the following Euler-Lagrange
equations

\begin{equation} \label{skalar}
(\partial_\mu+iqA_\mu)(\partial^\mu+iqA^\mu)\Phi+{\lambda \over 2}\Phi
(|\Phi|^2-{2m^2 \over \lambda})=0,
\end{equation}

\begin{equation} \label {wektor}
\partial_\mu F^{\mu\nu}=iq(\Phi^*\partial^\nu\Phi-\Phi\partial^\nu\Phi^*)
-2q^2A^\nu|\Phi|^2.
\end{equation}

The Higgs field $ \Phi(x) $ is complex valued, the star denotes the complex
conjugation. The signature  of metric of the space-time is $ (-+++) $.
The static Abrikosov-Nielsen-Olesen solution \cite{ANO} represents a straightlinear,
infinite vortex. The unit topological charge vortex lying along the z-axis
can be obtained from the equations (\ref{skalar}),(\ref{wektor}) by imposing
on them the axially symmetric Ansatz

\begin{eqnarray} \label{ansatz}
\Phi(\rho,\phi)  =  \sqrt{2m^2 \over \lambda} e^{i\theta}F(\rho), \nonumber \\
 A_0  =  0 ,\  A_3  =  0, \hbox{\quad \ \    } \nonumber \\
A_1  =  \sin(\theta){{\chi(\rho)-1}\over q\rho},  \\
A_2  =  -\cos(\theta){{\chi(\rho)-1}\over q\rho}. \nonumber  
\end{eqnarray}
Here  $\rho=\sqrt{(x^1)^2+(x^2)^2}$ and $\theta=\arctan(x^2/x^1)$ are
the standard polar coordinates  in the  $(x^1,x^2)$  plane. 

The Ansatz (\ref{ansatz}) together with the  rescaling of the $\rho$ variable

\begin{equation}
r=2m^2 \rho,
\end{equation}
leads to the second order differential equations for the fields $F(r)$ 
and $\chi(r)$

\begin{equation} \label{feqn} 
F^{\prime\prime} + {F^\prime \over r} - {\chi^2 \over r^2} F + {1 \over 2}(F-F^3)=0,
\end{equation}

\begin{equation} \label{heqn}
\chi^{\prime\prime} - {\chi^\prime  \over r} - \kappa^2 F^2 \chi = 0,
\end{equation}
where prime denotes  differrentiation with respect  to r 
and $ \kappa=\sqrt{2 q^2\over \lambda} $ is the only remaining  free parameter. 

To guarantee the vortex solution to be regular on $x^3$-axis 
and to have finite energy per unit of length in $ x^3 $ direction 
the above equations must be supplemented by the  boundary conditions

\begin{equation} \label{fbound}
F(0)=0, F(\infty)=1,
\end{equation}

\begin{equation} \label{hbound}
\chi(0)=1, \chi(\infty)=0.
\end{equation}

The asymptotic form of $\chi$ can be easily obtained from eqs. 
(\ref{heqn}),(\ref{hbound}) simplified by putting $ F=1 $. Thus we  get

\begin{equation} \label{hasym}
\chi_{asym}(r)=c_0 r K_1(\kappa r) \approx c_0 \sqrt{\pi r \over 2} \exp(-\kappa r),
\end{equation}
where $K_1$ is the modified Bessel function \cite{BESSEL} and  $c_0$ is a constant. 
The asymptotic behaviour of $F(r)$ strongly depends on $ \kappa $ \cite{TAUBES}.
For $\kappa < {1 \over 2}$ it is determined mainly by the term 
$ \chi^2 F/ r^2 $ and can be obtained from eqs.  (\ref{feqn}),(\ref{fbound})

\begin{equation} \label{fasym1}
F_{asym1} = 1-{{c_0^2 \pi} \over {\kappa(1-4 \kappa^2) r}} \exp(-2 \kappa r),
\end{equation}
while for  $\kappa > {1 \over 2}$ the Higgs term $ (F-F^3)/2 $ is more 
important. In this case eq. (\ref{feqn}) linearized in $ (F-1) $ gives 
the following result

\begin{equation} \label{fasym2}
F_{asym2} = 1 + c_1 K_0(r) \approx 1 + c_1 \sqrt{\pi \over 2 r} \exp(-r) .
\end{equation}
Here $ K_0 $ denotes the zero-order modified Bessel function \cite{BESSEL}
and $ c_1 $ is a constant.

The asymptotic forms of $ F $ and $ \chi $ were used in \cite{AH1} to get 
the approximate static vortex solution. In the neighbourhood of $ r=0 $ 
the fields $ F $ and $ \chi $ were approximated by low order polynomials 
obtained as power series solutions of eqs.(\ref{feqn}),(\ref{heqn}). 
These polynomials were smoothly matched with the appropriate asymptotics 
at some point $ r = r_0 $ i.e. the functions $ F(r) $ and $ \chi(r) $
were required to be continuous at $ r = r_0 $ together with their 
first derivatives

\begin{equation} \label{match1}
F_{poly}(r_0)=F_{asym}(r_0), \ F_{poly}^{\prime}(r_0)=F_{asym}^{\prime}(r_0)
\end{equation},

\begin{equation} \label{match2}
\chi_{poly}(r_0)=\chi_{asym}(r_0), \ \chi_{poly}^{\prime}(r_0)=\chi_{asym}^{\prime}(r_0).
\end{equation} 

In the simplest version proposed in \cite{AH1} the static vortex solution  was
approximated by the following formulae

\begin{equation}
F(r)=\left\{\begin{array}{ll}
f_1 r - {1 \over 3!} f_3 r^3 & \textrm{if $ r < r_0 $}\\
1 & \textrm{if $ r > r_0, $}
\end{array} \right.
\end{equation}
  
\begin{equation} \label {h6}
\chi(r)=\left\{\begin{array}{ll}
1 - {1 \over 2!} \chi_2 r^2 + {1 \over 4!} \chi_4 r^4 - {1 \over 6!}  \chi_6 r^6 & \textrm{if $ r < r_0 $}\\
c_0 r K_1(r) & \textrm{if $ r > r_0, $}
\end{array} \right.
\end{equation}
involving four constants $ r_0, f_1, \chi_2, c_0 $. These constants were fixed
by applying the matching conditions described above. The other constants are
given by recurrence relations (\ref{f5h6}) below.

In order to get more accurate solutions the formula for $ F $ in the region 
$ r > r_0 $ was replaced with a more subtle one

\begin{equation} \label{fap}
F=\sqrt{1-2\bigg({\chi \over r}\bigg)^2},
\end{equation}
obtained from eq. (\ref{feqn}) simplified by neglecting the terms 
with the derivatives of $ F $. In this case the polynomial approximation 
of $ F $ must be completed with the term proportional to $ r^5 $

\begin{equation} \label {f5}
F=f_1 r - {1 \over 3!} f_3 r^3 + {1 \over 5!} f_5 r^5,
\end{equation}   
while the formulae for $ \chi $ remain unchanged although 
the values of the particular parameters are different. The equations
(\ref{feqn}),(\ref{heqn}) lead to the following recurrence relations
for coefficients of the polynomials

\begin{eqnarray} \label {f5h6}
f_3 &=& {3 \over 4}({1 \over 2}+\chi_2)f_1, \nonumber \\
\chi_4 &=& 3 \kappa^2 f_1^2, \nonumber \\
f_5 &=& {5 \over 6}({1 \over 2}+\chi_2) f_3 + {5 \over 2}({1 \over 6}  \chi_4 +
{1 \over 2} \chi_2^2 + f_1^2) f_1, \\
\chi_6 &=& 5 \kappa^2 f_1 (2 f_3+3 \chi_2 f_1). \nonumber 
\end{eqnarray}

In Fig. (1a)-(1c) we  have compared the described above approximation 
of the Higgs  field $ F $ with its numerical values obtained by applying 
standard algorithms for stiff differential equations \cite{NR}. 
As we are interested mainly in small values of $ \kappa $ we have limited  
ourselves to $ \kappa=0.02 $, $ \kappa=0.1 $ and $ \kappa=0.2 $. Since 
the values of the field  $ \chi $ obtained from the numerical 
computations and approximate formula differ very slightly we have plotted 
their differences ( Fig. (2) ) and the numerical values themselves 
( Fig. (3) ). The numerical values of the free parameters $ f_1,\chi_2,
r_0,c_0 $ are given below. 
\vspace{0.5cm}

\hspace{3.0cm} ($ F_{asym}=1 $)

\vspace{0.2cm}

\begin{tabular}{|c|c|c|c|c|}
 \hline
 $\kappa$ & $f_1$ & $h_2$ & $r_0$ & $c_0$ \\
 \hline
 0.02 & 0.6505427 & 0.00157722 & 2.305767 & 0.020004 \\
 0.1 & 0.6646855 & 0.02362296 & 2.256706 & 0.1005027 \\
 0.2 & 0.6929167 & 0.06904716 & 2.164762 & 0.2040623 \\
 \hline
\end{tabular}
\vspace{0.5cm}

\hspace{2.0cm} ($ F_{asym}=\sqrt{1-2(\chi /r)^2} $)

\vspace{0.2cm}

\begin{tabular}{|c|c|c|c|c|}
 \hline
 $\kappa$ & $f_1$ & $h_2$ & $r_0$ & $c_0$ \\
 \hline
 0.02 & 0.4285536 & 0.001431718 & 2.106883 & 0.0200066 \\
 0.1 & 0.443162 & 0.02028046 & 2.068759 & 0.1007802 \\
 0.2 & 0.469365 & 0.05754442 & 2.000099 & 0.2056864 \\
 \hline
\end{tabular}
\vspace{0.5cm}

Let us note that the approximate formula for the function $ \chi $ 
is quite good while the approximation of $ F $ is much worse. 
This is the price for simplicity of the analytical expressions.
The field $ F $ tends very quickly to its asymptotic form 
and such behaviour can be hardly described by simple analytical formula.

\section{Improved approximate solutions}
\vspace{.5cm}
\indent
The main defect of the approximate formulae considered in the previous section 
is the behaviour of the field $ F $ in the region of intermediate values of $ r $ 
particularly in the neighbourhood of the matching point $ r_0 $. One can try to 
improve that approximation by using the higher order polynomial solution 
for the functions $ F $ and $ \chi $ in the interval $ (0,r_0) $. 
However the practical effect of such improvement seems to be rather small. 
A better accuracy can be reached by changing the approximation for the function 
$ F $ in the region $ (r_0, \infty) $. The more accurate asymptotics \cite{AH1} is 
given by

\begin{equation} \label{fasymfull}
F_{asym}=\sqrt{1-2\bigg({\chi_{asym} \over r}\bigg)^2} + c_1 K_0(r).
\end{equation}

This formula involves a new parameter $ c_1 $ and an extra condition is neccessary
to determine it. Therefore we have used an additional matching condition ensuring 
the continuity of the second order derivative of $ F $ at $ r=r_0 $

\begin{equation}
F_{poly}^{\prime\prime}(r_0)=F_{asym}^{\prime\prime}(r_0).
\end{equation}

We were able to satisfy the matching conditions if the polynomials approximating 
the functions $ F $ and $ \chi $ were of order fifteen and fourteen or nineteen
and eighteen, respectively. We do not present them as their forms are very
complicated and the numerical results not excellent as is shown in Fig. (4) and (5).

There is also another simple posssibility to determine the values of five 
parameters $ c_0, c_1, f_1, h_2, r_0 $. One can solve the four matching 
conditions (\ref{match1}),(\ref{match2}) for the fixed value of the radius $ r_0 $ 
and repeat this procedure for several values of $ r_0 $ in some interval.
Thus we have the four parameters $ c_0, c_1, f_1, h_2 $ as the numerical 
functions of $ r_0 $ . The last step was to compare the approximations 
for $ F $ an $ \chi $ obtained this way with numerical calculations and 
fix the value of $ r_0 $ which gives the best fitting. It turned out that 
in this case it was enough to approximate $ F $ and $ \chi $ by polynomials of 
order five and four respectively, see eqs. (\ref {h6}), (\ref {f5}), 
(\ref {f5h6}) with neglected $ \chi_6-terms $. 
The differences between approximated and numerical values of $ F $ and $ \chi $ 
are presented in Fig. (6) and (7). The numerical values of the  parameters 
$ f_1,\chi_2,c_0,c_1 $ are given in the following table

\vspace{0.5cm}

\begin{tabular}{|c|c|c|c|c|}
 \hline
 $\kappa$ & $f_1$ & $h_2$ & $c_0$ & $c_1$ \\
 \hline
 0.02 & 0.3811035 & 0.001427667 & 0.0200068 & 1.368773 \\
 0.1 & 0.3885089 & 0.02025245 & 0.1007819 & 1.530144 \\
 0.2 & 0.3998954 & 0.05794554 & 0.2052965 & 1.851151 \\
 \hline
\end{tabular}

\vspace{0.5cm}
\noindent
while $ r_0=2.5 $.
The vortex solutions considered so far were obtained by smooth matching 
of some polynomials approximating the vortex core with asymptotic formulae 
valid in the outer region. More accurate approximations can be obtained by 
dividing the whole area into more pieces and approximating the solution in 
each sector separately. We have chosen $ k+1 $ points $ 0<r_0<r_1<...<r_k $. 
The central part of the vortex in the interval $ (0,r_0) $ was approximated by
 
\begin{equation} \label{f0full}
F = f_1 r - {1 \over 3!} f_3 r^3 + {1 \over 5!} f_5 r^5 + ... + 
{(-1)^n \over (2n+1)!} f_{2n+1} r^{2n_0+1},
\end{equation}

\begin{equation} \label{h0full}
\chi= 1 - { 1 \over 2! } \chi_2 r^2 + {1 \over 4!} \chi_4 r^4 + ... +
{(-1)^n \over (2n)!} \chi_{2n} r^{2n_0},
\end{equation}
while in $ (r_j,r_{j+1}) $ for $ j=0,1,2,..k-1 $ we have used the truncated 
Taylor series expansions
\begin{equation} \label{f1full}
F = \widetilde{f}_{0j} + \widetilde{f}_{1j} (r-r_j) + 
{1 \over 2!} \widetilde{f}_{2j} (r-r_j)^2 + ... + 
{1 \over n!} \widetilde{f}_{nj} (r-r_j)^{n_j},
\end{equation}

\begin{equation} \label{h1full}
{\chi}= \widetilde{\chi}_{0j}+ \widetilde{\chi}_{1j} (r-r_j) + 
{1 \over 2!} \widetilde{\chi}_{2j} (r-r_j)^2 + ... +
{1 \over n!} \widetilde{\chi}_{nj} (r-r_j)^{n_j}.
\end{equation}
In the region $ (r_k,\infty) $ the previous asymptotic formulae (\ref{hasym}),
(\ref{fasymfull}) were applied. We have required the functions $ F $ and 
$ \chi $ together with their first derivatives to be continuous in the matching 
points $ r_0,r_1,...,r_k $. These conditions and eqs. (\ref{feqn}),(\ref{heqn})
are enough to determine all the coefficients 
$ f_j,\chi_j,f_{ij},\chi_{ij},c_0,c_1 $. 
Let us note that passing from $ r=0 $ to $ r=r_k $ resembles the process of
analytic continuation and is rather easy to perform. The main difficulty is
to bind these solutions with their appropriate asymptotics.   

The above formulae were applied in two ways. In equations (\ref{f1full}), 
(\ref{h1full}) we have firstly put the expansion order to 4 ($ n_0=2,n_1=4 $)
 and have divided the whole region of $ r $ into three pieces ($ r_0=2,r_1=3 $).
In the region $ (0,r_0) $ the formulae (\ref{f0full}),(\ref{h0full}) for 
$ n_0=2 $ reduce to (\ref {h6}), 
(\ref {f5}), (\ref {f5h6}) with neglected $ \chi_6-terms $ while 
in the interval $ (r_0,r_1) $
the following recurrence relations are valid in eqs. 
(\ref {f1full}),(\ref {h1full})
\begin{eqnarray}
\widetilde{f}_{2}&=&{-\widetilde{f}_{0} \over 2}+
{\widetilde{f}_{0}^3 \over 2}-{\widetilde{f}_{1} \over r_0}+
{ \widetilde{f}_{0} \widetilde{\chi}_{0}^2 \over r_0^2}, \nonumber \\
\widetilde{\chi}_{2}&=&{\kappa^2 \widetilde{f}_{0}^2 \widetilde{\chi}_{0}}+
{\widetilde{\chi}_{1} \over r_0}, \nonumber \\
\widetilde{f}_{3}&=&{\widetilde{f}_{1} \over  r_0^2}+
{2 \widetilde{f}_{0} \widetilde{\chi}_{0} \widetilde{\chi}_{1} \over r_0^2}
+{\widetilde{\chi}_{0}^2 \widetilde{f}_{1} \over r_0^2}-
{\widetilde{f}_{2} \over r_0}-{\widetilde{f}_{1} \over 2}
+{3 \widetilde{f}_{0}^2 \widetilde{f}_{1} \over 2}-
{2 \widetilde{f}_{0} \widetilde{\chi}_{0}^2 \over r_0^3}, \nonumber \\
\widetilde{\chi}_{3}&=&-{\widetilde{\chi}_{1} \over r_0^2}+
{\kappa^2 \widetilde{f}_{0}^2 \widetilde{\chi}_{1}}
+{2 \kappa^2  f_{0} \widetilde{f}_{1} \widetilde{\chi}_{0}}+
{\widetilde{\chi}_{2} \over r_0}, \nonumber \\
\widetilde{f}_{4}&=&-{2 \widetilde{f}_{1} \over r_0^3}+
{6 \widetilde{f}_{0} \widetilde{\chi}_{0}^2 \over r_0^4}
-{\widetilde{f}_{3} \over r_0}-{f_{2} \over 2}+
{ 3 \widetilde{f}_{2} \widetilde{f}_{0}^2 \over 2}
+{ 3 \widetilde{f}_{1}^2 \widetilde{f}_{0} } \nonumber \\
&+&{2 \widetilde{f}_{0} \widetilde{\chi}_{0} \widetilde{\chi}_{2} \over r_0^2}+
{2 \widetilde{f}_{0} \widetilde{\chi}_{1}^2 \over r_0^2}
+{ \widetilde{f}_{2} \widetilde{\chi}_{0}^2 \over r_0^2}+
{4 \widetilde{f}_{1} \widetilde{\chi}_{0} \widetilde{\chi}_{1} \over r_0^2}
+{2 \widetilde{f}_{2} \over r_0^2} \nonumber \\
&-&{8 \widetilde{f}_{0} \widetilde{\chi}_{0} \widetilde{\chi}_{1} \over r_0^3}
-{4 \widetilde{f}_{1} \widetilde{\chi}_{0}^2 \over r_0^3}, \nonumber \\
\widetilde{\chi}_{4}&=&-{2 \widetilde{\chi}_{2} \over r_0^2}+
{2 \widetilde{\chi}_{1} \over r_0^3}
+{\kappa^2 \widetilde{f}_{0}^2 \widetilde{\chi}_{2}}+
{4 \kappa^2 \widetilde{f}_{0} \widetilde{f}_{1} \widetilde{\chi}_{1}} \nonumber \\
&+&{2 \kappa^2  \widetilde{f}_{0} \widetilde{\chi}_{0} \widetilde{f}_{2}}
+{2 \kappa^2 \widetilde{\chi}_{0} \widetilde{f}_{1}^2 }+
{\widetilde{\chi}_{3} \over r_0}, \nonumber 
\end{eqnarray}
where the second index $ (j=0) $ is omited for simplicity.
The numerical values of the remaining parameters are found from the matching 
conditions and they are given in the
following two tables 

\vspace{0.5cm}

\begin{tabular}{|c|c|c|c|c|}
 \hline
 $\kappa$ & $f_1$ & $h_2$ & $c_0$ & $c_1$ \\
 \hline
 0.02 & 0.3980053 & 0.1392637 & 0.02000953 & 1.273493 \\
 0.1 & 0.4078238 & 0.01947355 & 0.1011135 & 1.34181 \\
 0.2 & 0.4238514 & 0.05537332 & 0.2079306 & 1.479463 \\
 \hline
\end{tabular}

\vspace{0.5cm}

\begin{tabular}{|c|c|c|c|c|}
 \hline
 $\kappa$ & $\widetilde{f}_{0}$ & $\widetilde{f}_{1}$ & $\widetilde{\chi}_{0}$ & $\widetilde{\chi}_{1}$ \\
 \hline
 0.02 & 0.6551700 & 0.2454606 & 0.9973415 & -0.002531821 \\
 0.1 & 0.6676333 & 0.2496415 & 0.9643793 & -0.03229428 \\
 0.2 & 0.6863076 & 0.2557591 & 0.9036254 & -0.08200263 \\
 \hline
\end{tabular}
\vspace{0.5cm}

\noindent
The rather simple but quite accurate analytical approximation 
of the static vortex solution was obtained this way, as is presented in 
Fig. (8) and (9).
   
In the second case we have tried to get the possibly most accurate numerical results.
Therefore we have put large $ n_0=9,n_j=10 $ and very small value of 
$ r_{j+1}-r_j=0.01 $ for $ r_0=1,r_k=20 $ . 
The corresponding results are shown in Fig. (10) and (11).

\section{Remarks}
\indent
In the present paper we have considered several formulae approximating 
the vortex solution in the Abelian Higgs model. We started with simple 
analytical formulae presented in \cite{AH1} and compared them with numerical 
computations. It turned out that the approximation of the Higgs
field in the neighbourhood of the vortex core is rather rough and should 
be improved to get more accurate results. We have tried several methods 
to reach this goal.

First of all we have changed the formula describing the Higgs field 
in the outer region. This formula should not be interpreted as the better 
asymptotics only. Perhaps more important is the fact that this expression 
involves a new free parameter which let us improve the whole algorithm. 
We have used this possibility in several ways. At this point it is worth
noting that the relative error of the function $ F-1 $ with $ F $ given by
the approximate formula (\ref{fap}) 
does not tend to zero for large r. This can be easily seen by comparing 
eqs. (\ref{hasym}), (\ref{fasym1}) and (\ref{fap}).

Our first trial to improve the accuracy of the approximation 
was the algorithm with an  additional 
matching condition ensuring the continuity of the second  derivative of the
Higgs field. The next possibility we have tried was to solve the matching 
conditions in some fixed point and repeat this step several times in 
different points to choose finally the best matching point on the basis 
of numerical results.

At last we have modified the algorithm by solving the equations of motion 
approximatelly as the truncations of the Taylor series expansions around 
an arbitrary point. We have used these solutions in the manner resembling 
the process of analytical continuation. This way we have obtained both: 
our best numerical approximations of the static vortex solutions and quite 
simple but accurate analytical formulae generalizing those from \cite{AH1}. 
It was possible  because the final version of the algorithm turned out 
to be very flexible and could be applied to reach apparently different 
purposes: analytical simplicity of expressions and numerical accuracy
of computer calculations.    

\vspace{1.cm}
\noindent
{\bf Acknowledgements.}\\
This work was supported in part by KBN grant No. 2 P03B 095 13.

\pagebreak

\pagebreak
\section{Figures}
\noindent
Fig. 1a. The approximate and numerical values of the field  $ F, \kappa=0.05 $. \\
\\
Fig. 1b. The approximate and numerical values of the field  $ F, \kappa=0.1 $. \\
\\
Fig. 1c. The approximate and numerical values of the field  $ F, \kappa=0.2 $. \\
\\
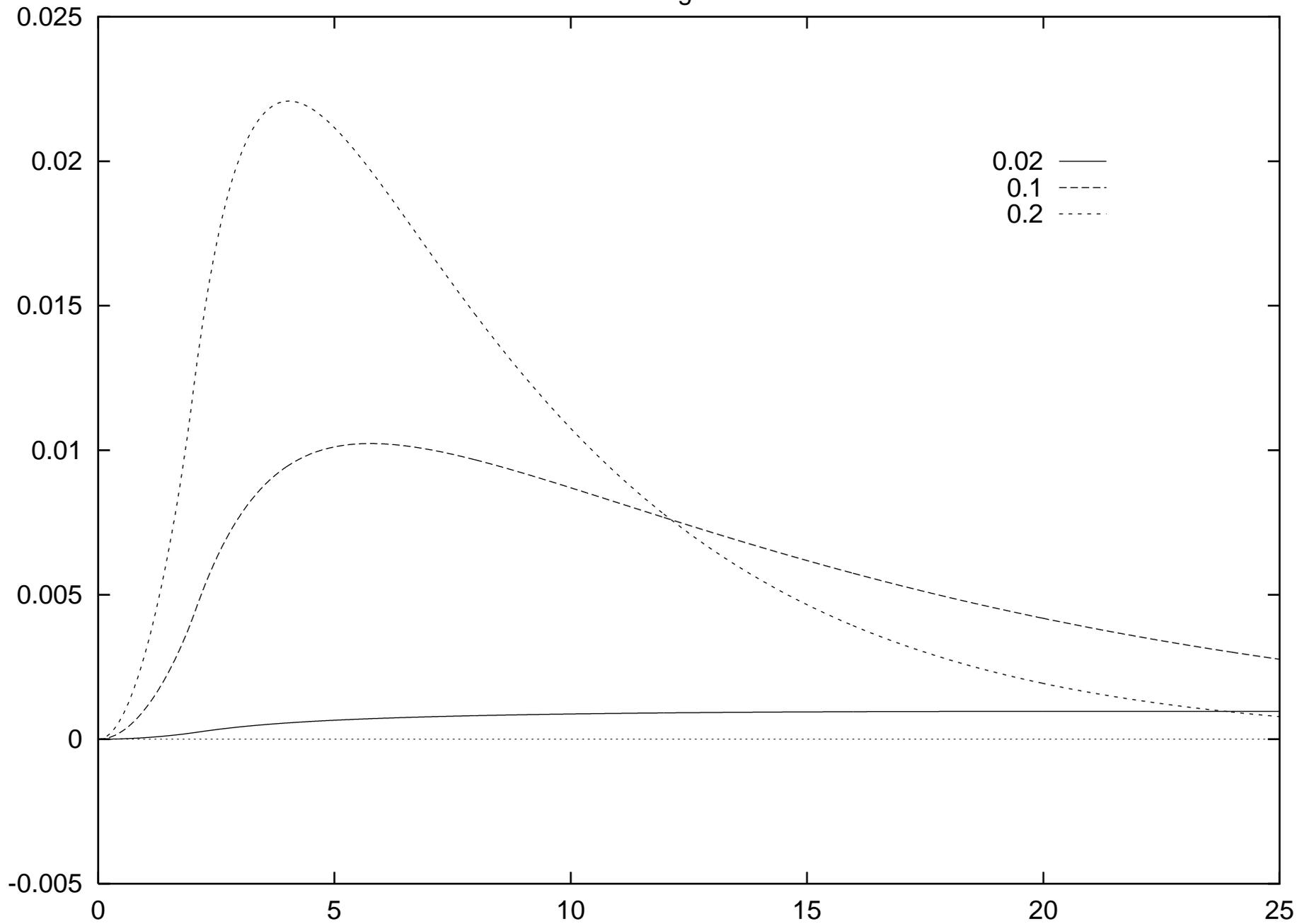
Fig. 2. The differences between approximate and numerical values  of the \\ 
\hspace*{1.1cm} field $ \chi $ for $ \kappa=0.05, 0.1, 0.2 $. \\
\\
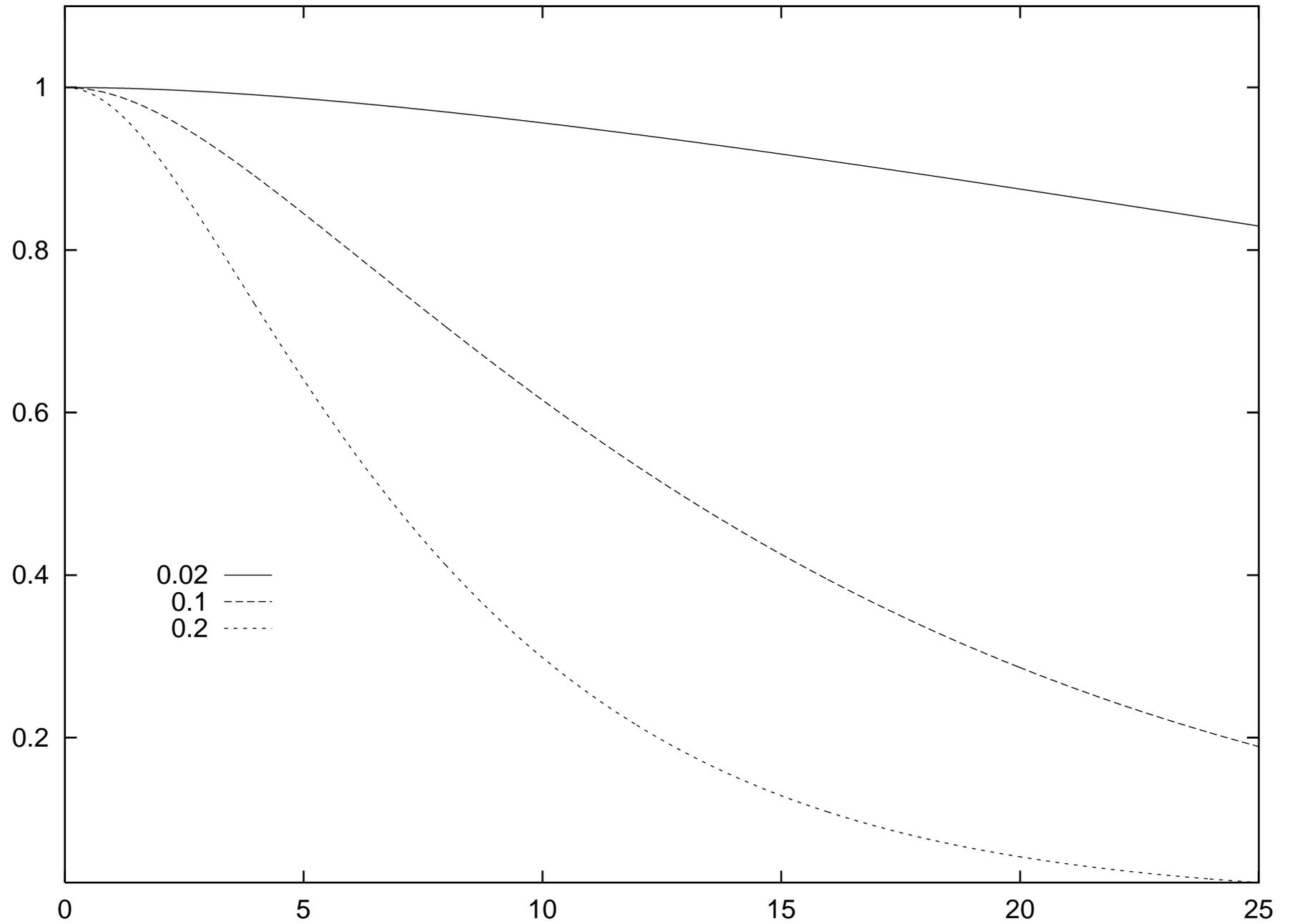
Fig. 3. The numerical values of the field  $ \chi $ for 
$ \kappa=0.05, 0.1, 0.2 $. \\
\\
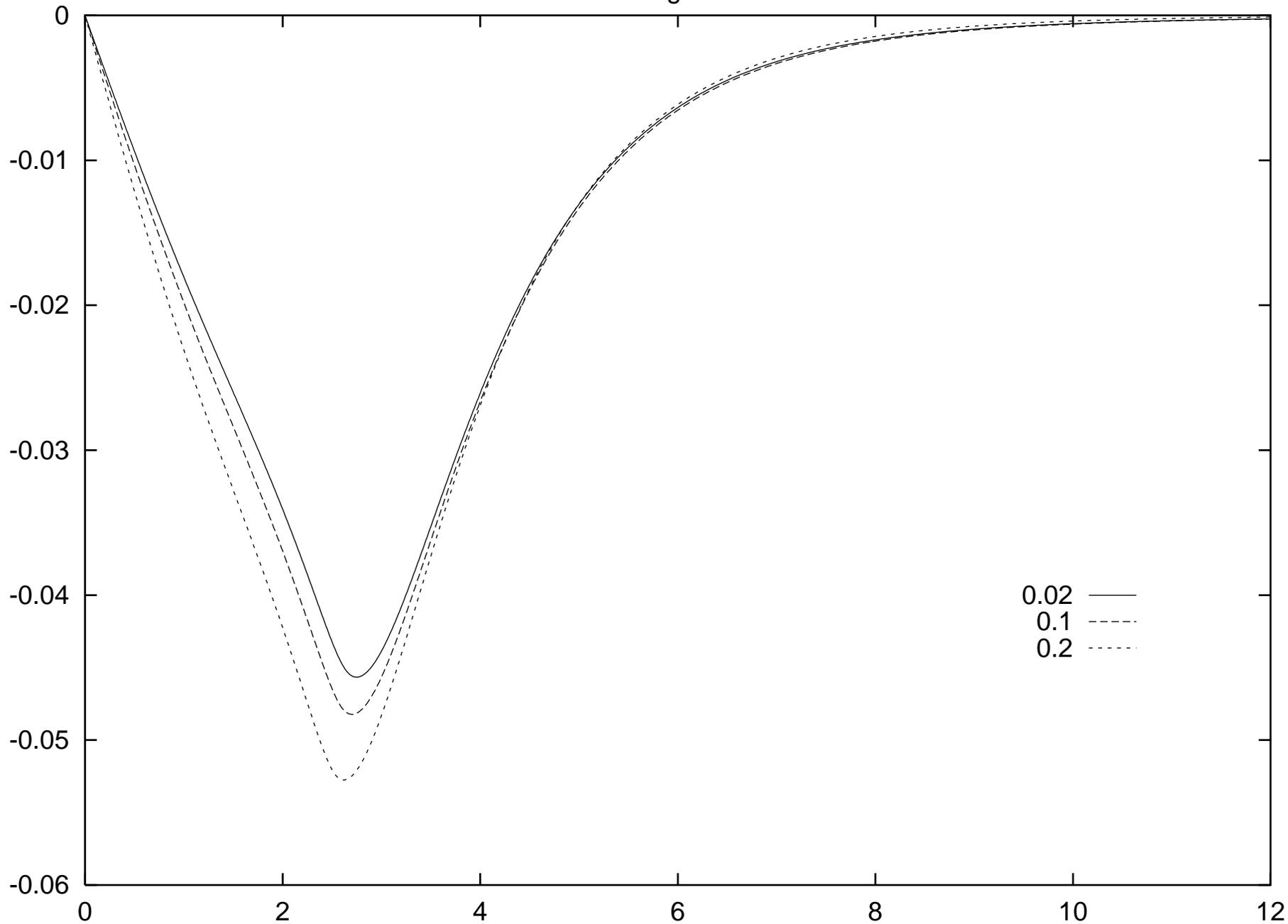
Fig. 4. The differences between approximate and numerical values of the \\
\hspace*{1.1cm} field $ F $ for $ \kappa=0.05, 0.1, 0.2 $ 
(second order derivative method). \\
\\
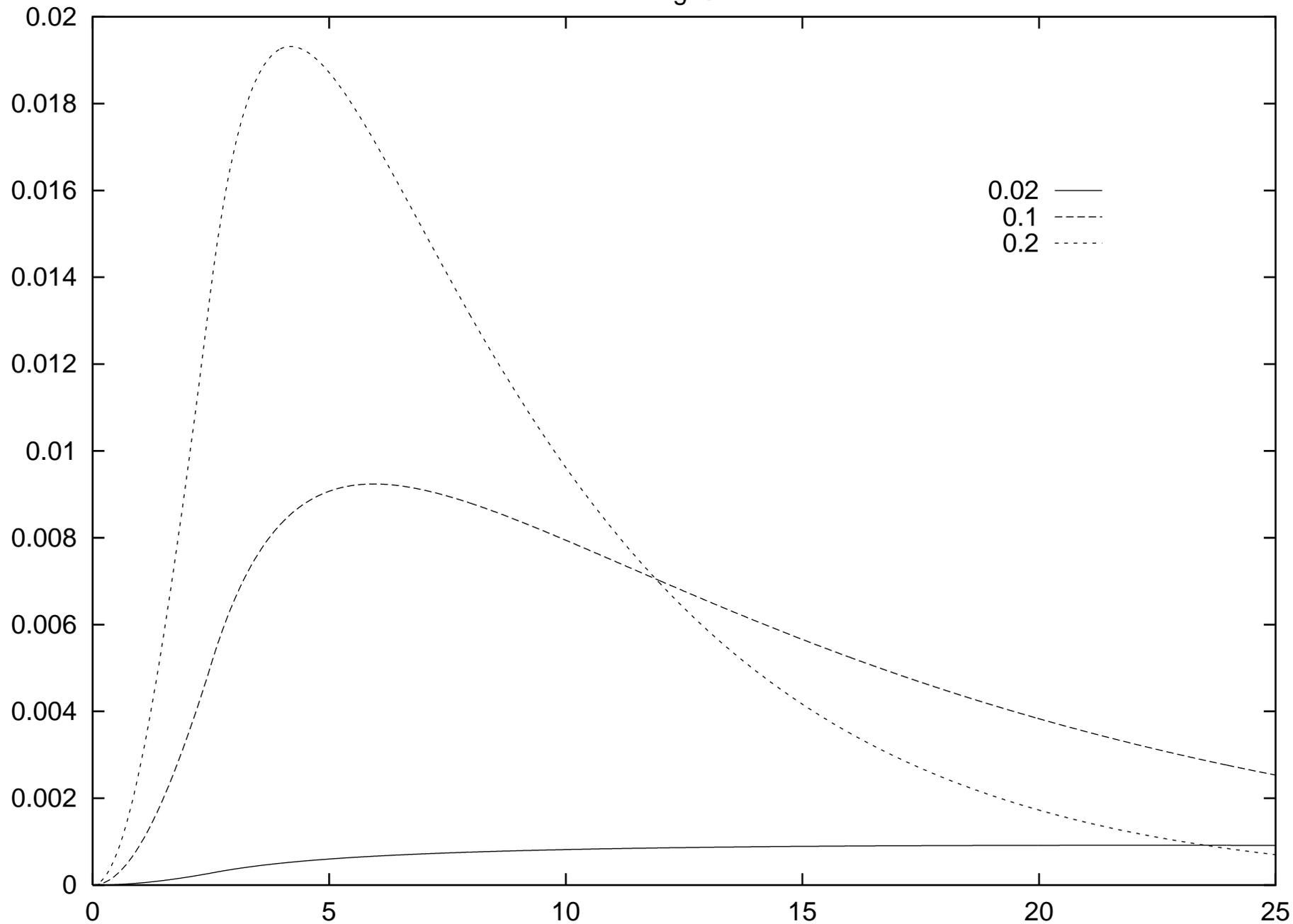
Fig. 5. The differences between approximate and numerical values of the \\
\hspace*{1.1cm} field $ \chi $ for $ \kappa=0.05, 0.1, 0.2 $ 
(second order derivative method). \\
\\
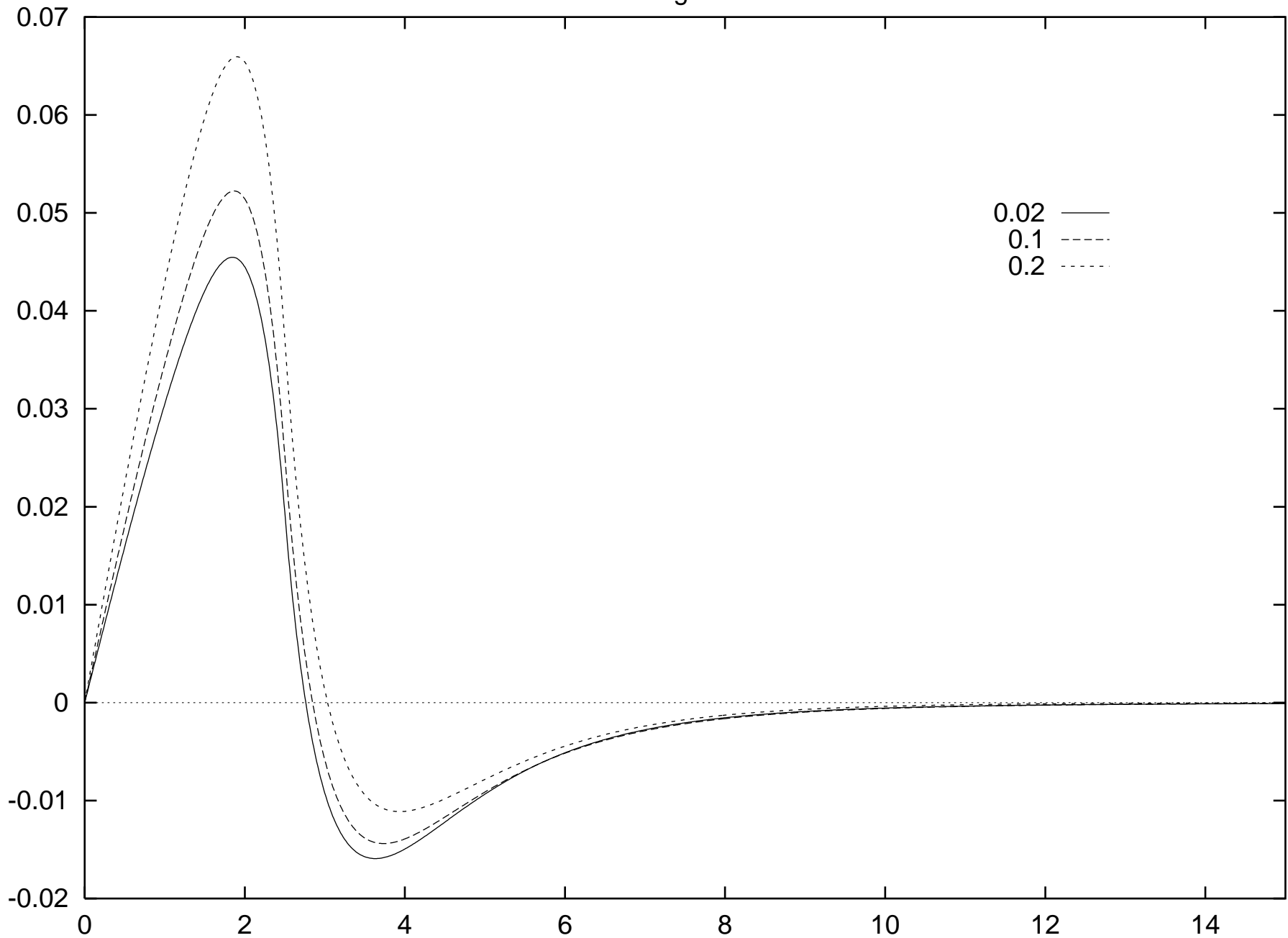
Fig. 6. The differences between approximate and numerical values of the \\
\hspace*{1.1cm} field $ F $ for $ \kappa=0.05, 0.1, 0.2 $ 
(fixed matching point algorithm). \\
\\
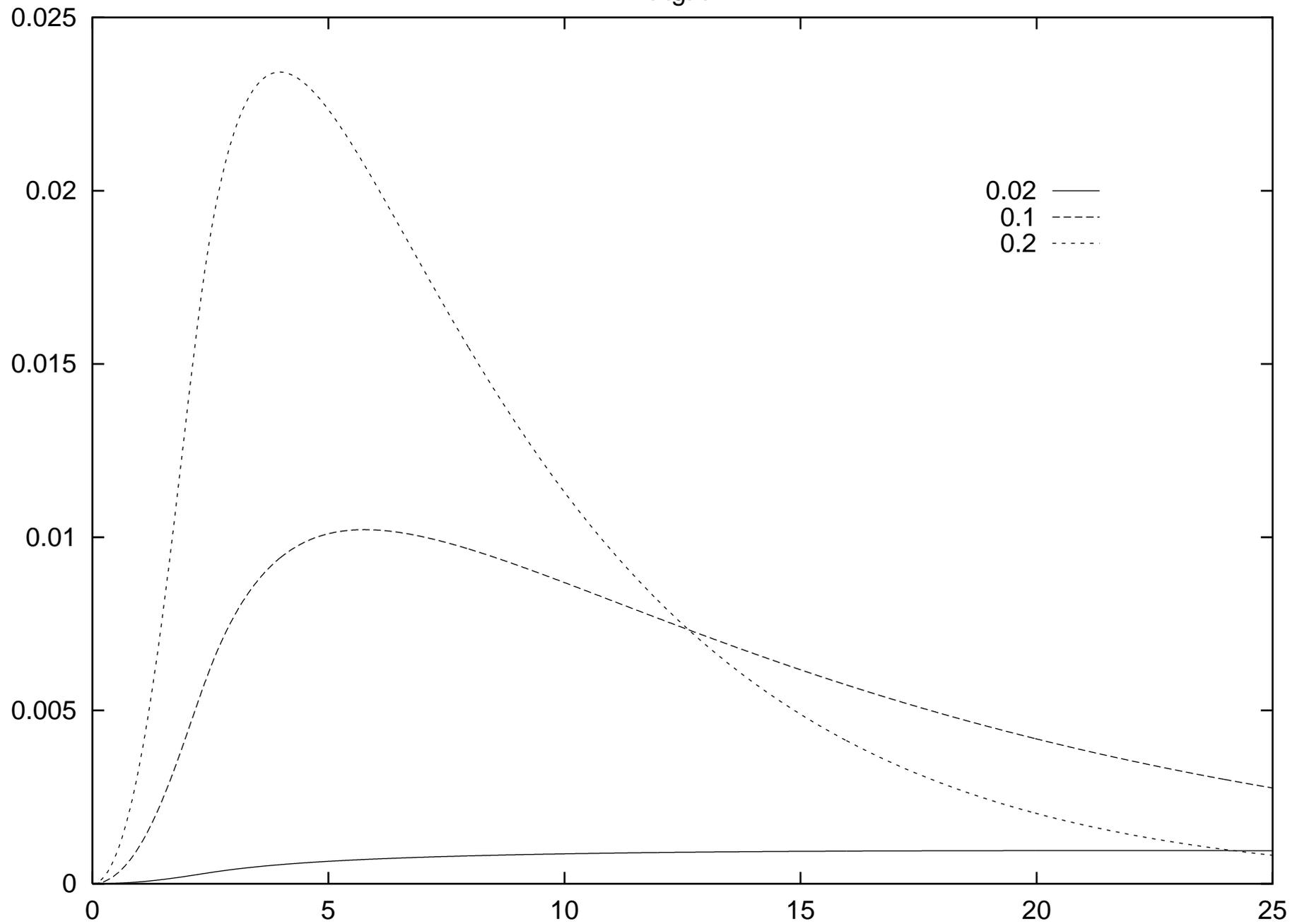
Fig. 7. The differences between approximate and numerical values of the \\
\hspace*{1.1cm} field $ \chi $ for $ \kappa=0.05, 0.1, 0.2 $ 
(fixed matching point algorithm). \\
\\
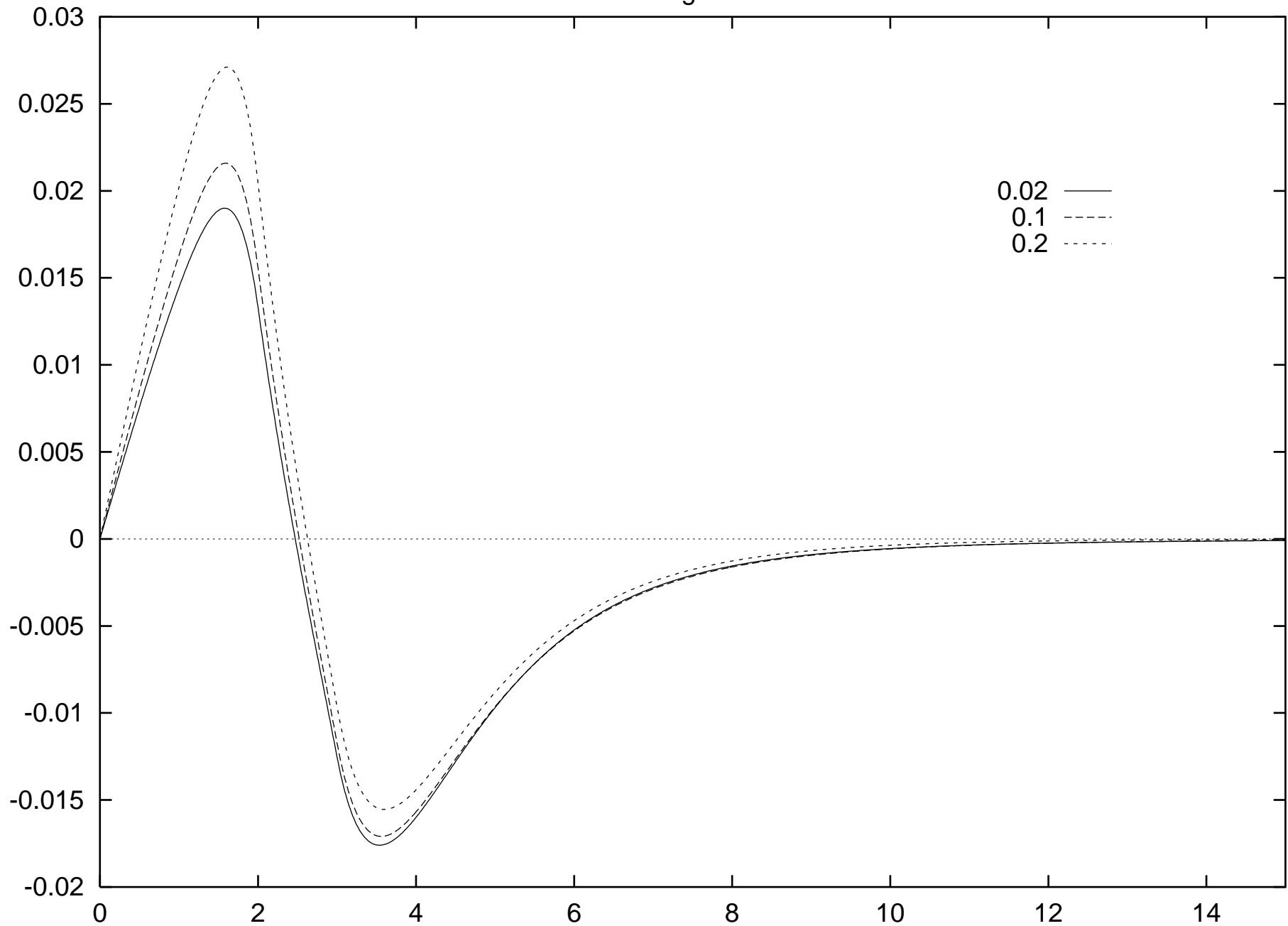
Fig. 8. The differences between approximate and numerical values of the \\
\hspace*{1.1cm} field $ F $ for $ \kappa=0.05, 0.1, 0.2 $ 
(two matching points method). \\
\\
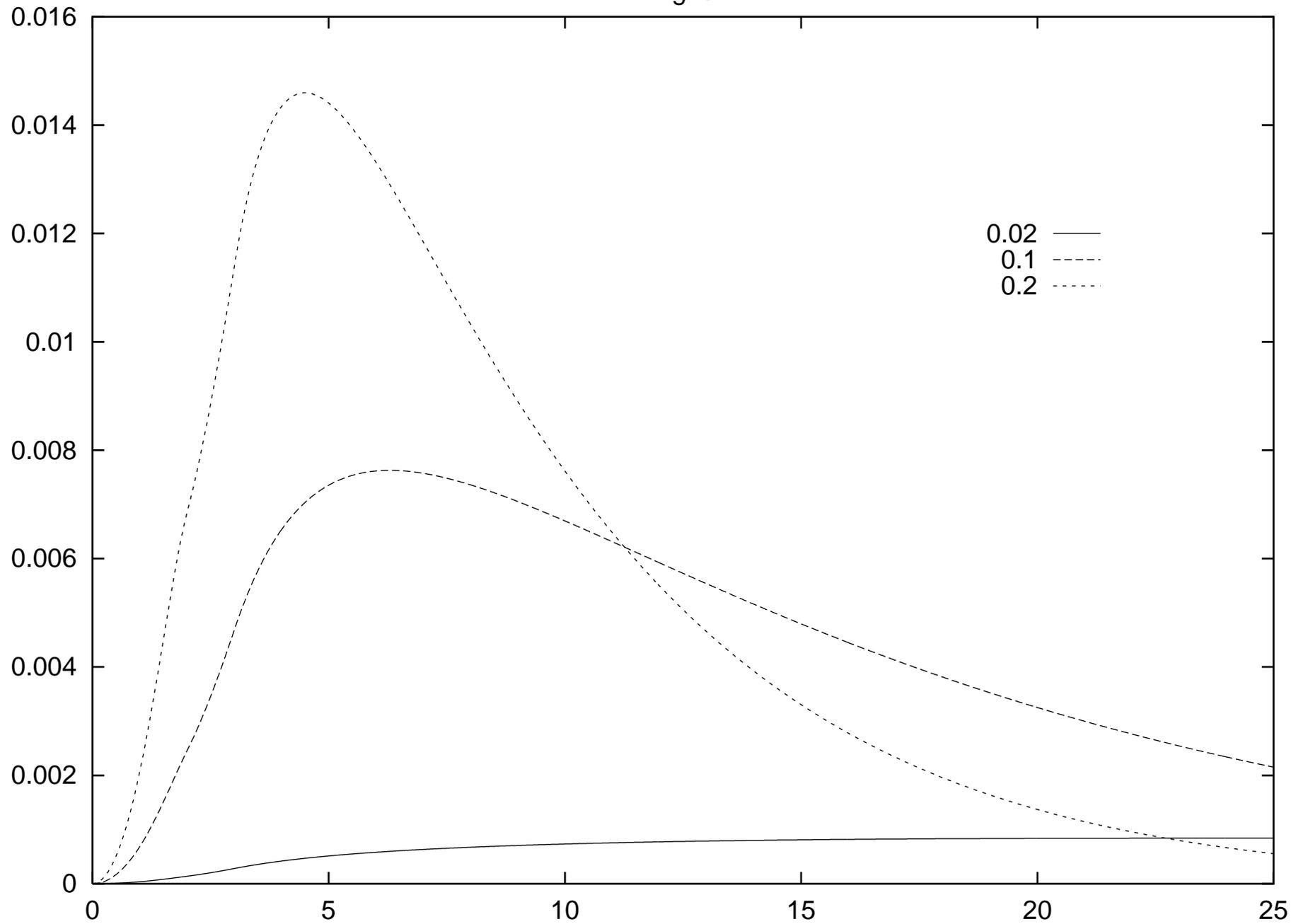
Fig. 9. The differences between approximate and numerical values of the \\
\hspace*{1.1cm} field $ \chi $ for $ \kappa=0.05, 0.1, 0.2 $ 
(two matching points method). \\
\\
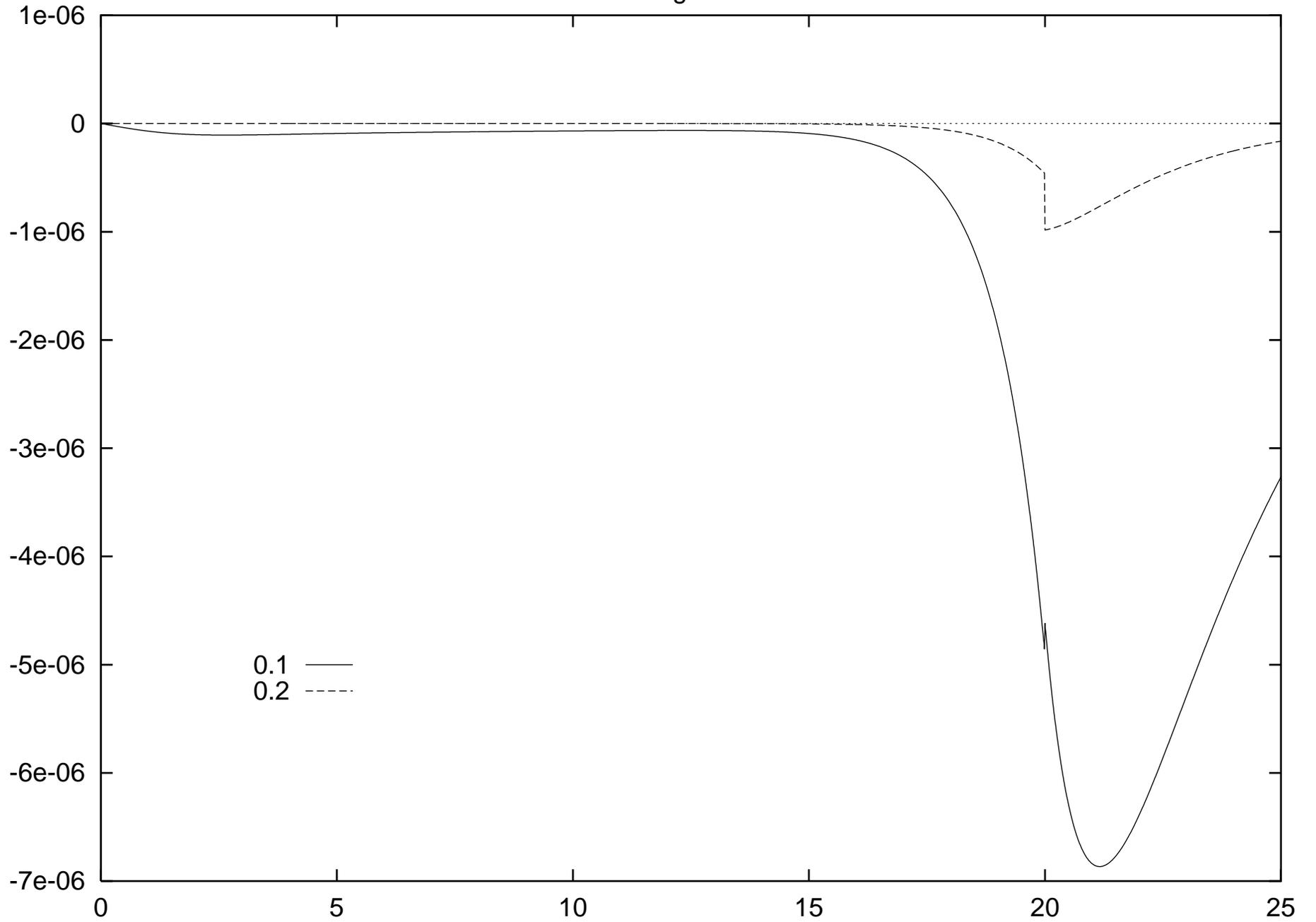
Fig. 10. The differences between approximate and numerical values of the \\
\hspace*{1.1cm} field $ F $ for $ \kappa=0.05, 0.1, 0.2 $ 
(analytic continuation alghorithm). \\
\\
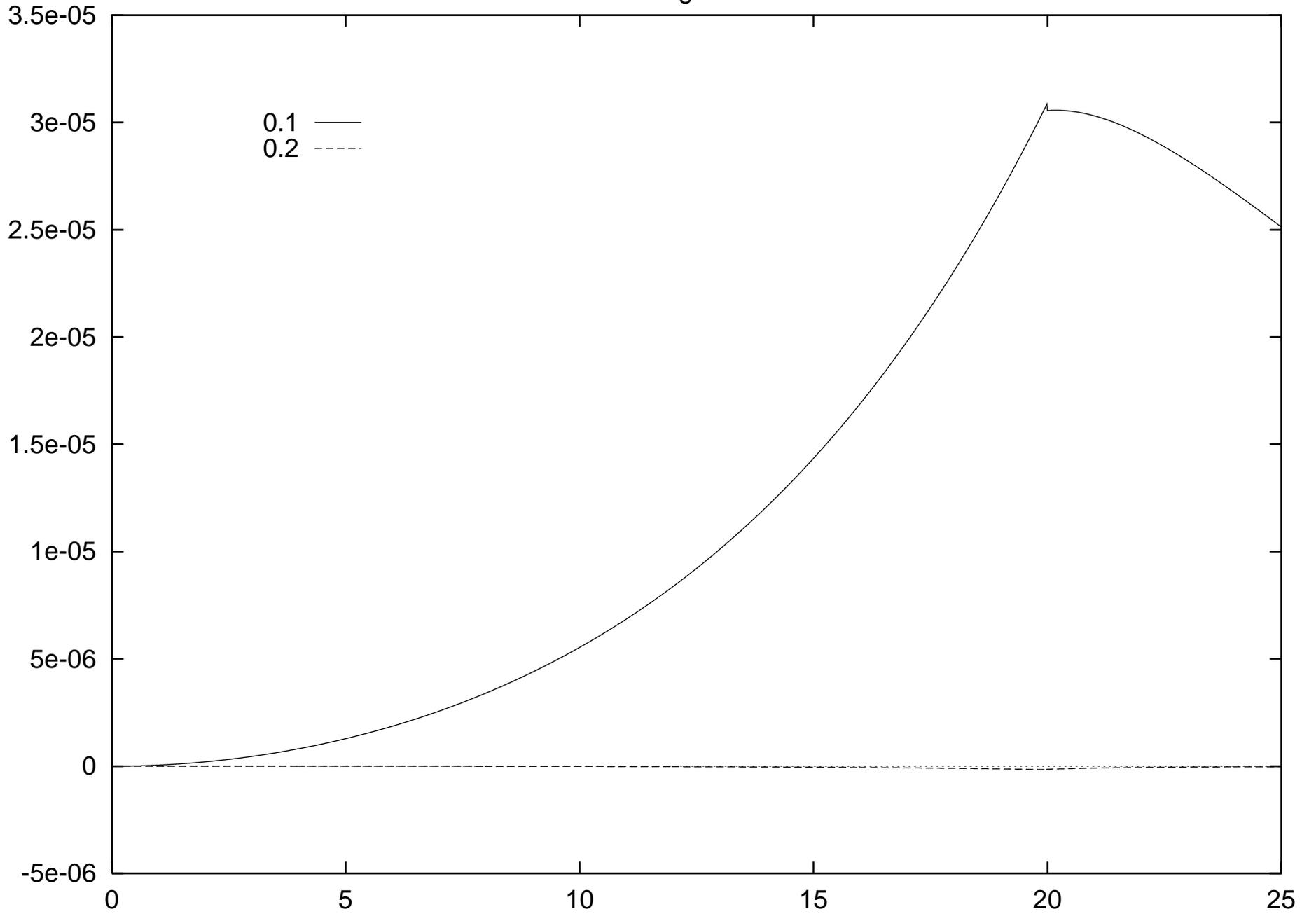
Fig. 11. The differences between approximate and numerical values of the \\
\hspace*{1.1cm} field $ \chi $ for $ \kappa=0.05, 0.1, 0.2 $ 
(analytic continuation alghorithm). 

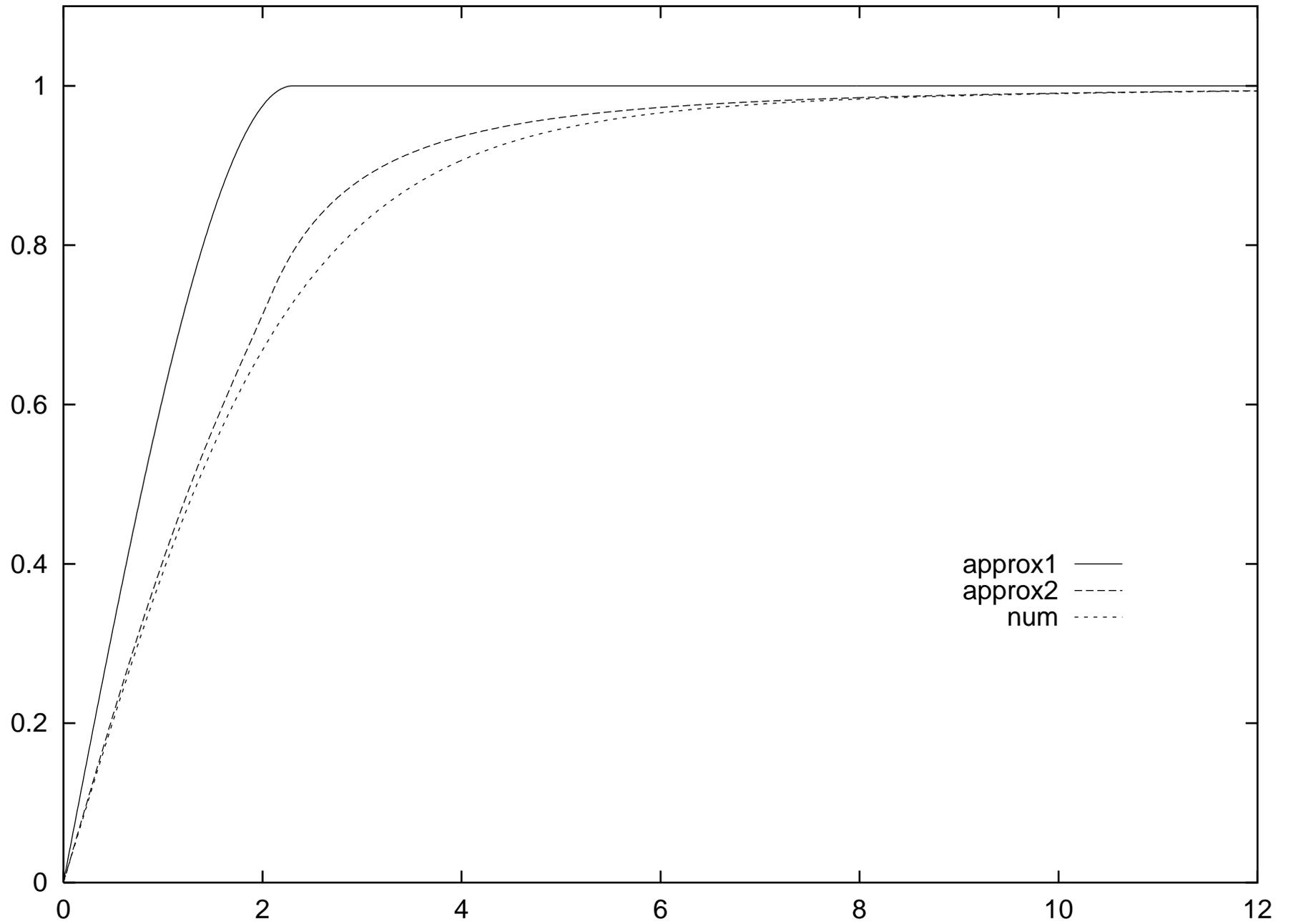
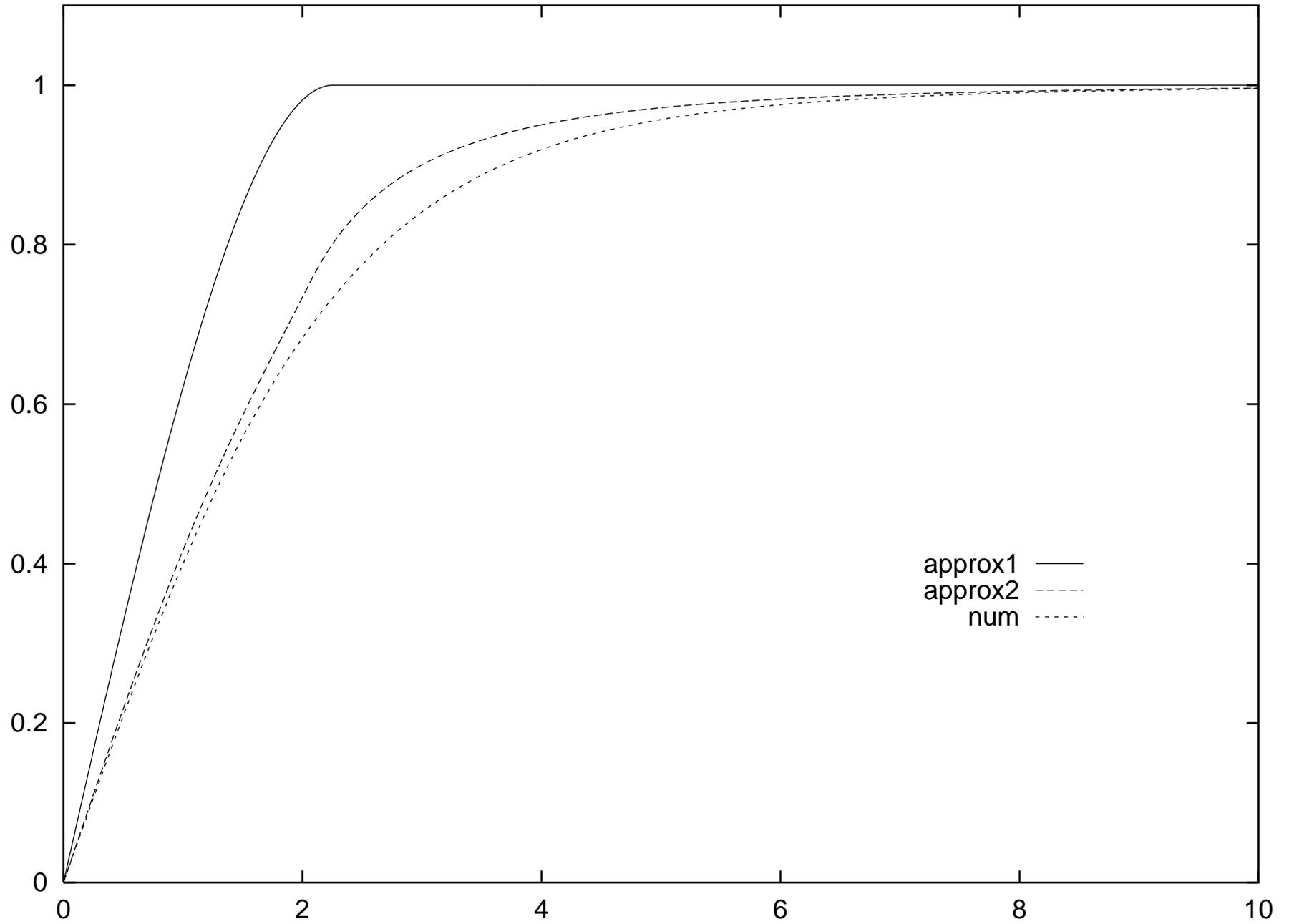
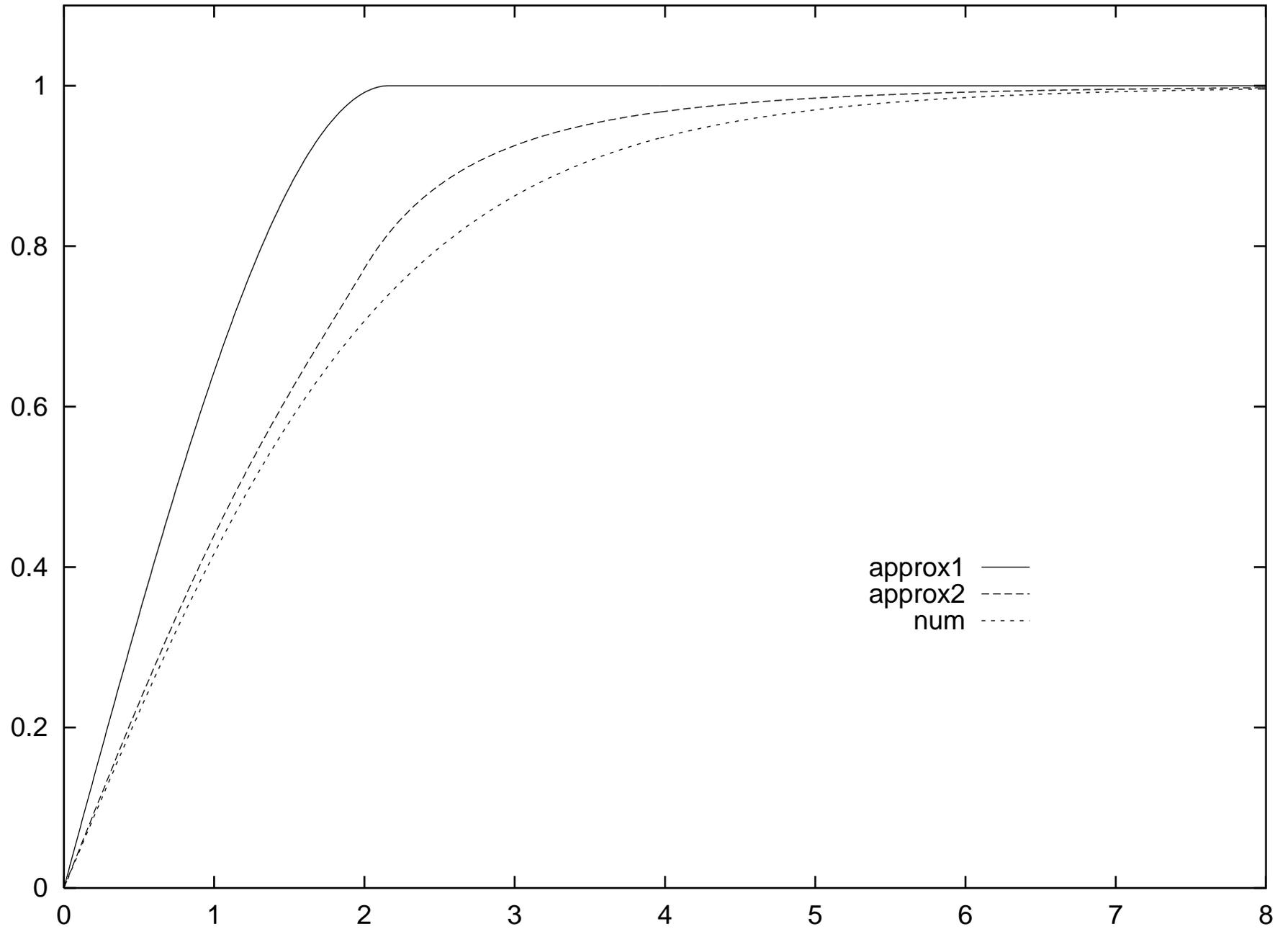
\end{document}